\documentclass[aps,prl,superscriptaddress,twocolumn,amsmath,amssymb,showpacs]{revtex4}
\usepackage{graphicx}% Include figure files
\usepackage{dcolumn}% Align table columns on decimal point
\usepackage{bm}% bold math
\usepackage{textcomp}% bold math

\begin{document}

\title{Optimizing the search for resources by sharing information: Mongolian gazelles as a case study.}

\author{Ricardo Mart\'inez-Garc\'ia}
\affiliation{IFISC, Instituto de F\'isica Interdisciplinar y Sistemas Complejos (CSIC-UIB),
  E-07122 Palma de Mallorca, Spain}

\author{Justin M. Calabrese}
\affiliation{Conservation Ecology Center,
Smithsonian Conservation Biology Institute,
National Zoological Park,
1500 Remount Rd., Front Royal, VA 22630, USA}

\author{Thomas Mueller}
\affiliation{Conservation Ecology Center,
Smithsonian Conservation Biology Institute,
National Zoological Park,
1500 Remount Rd., Front Royal, VA 22630, USA}
\affiliation{Behaviour, Ecology, Evolution, and systematics Program, University of Maryland, College Park, MD 20742, USA.}

\author{Kirk A. Olson}
\affiliation{Conservation Ecology Center,
Smithsonian Conservation Biology Institute,
National Zoological Park,
1500 Remount Rd., Front Royal, VA 22630, USA}

\author{Crist\'obal L\'opez} 
\affiliation{IFISC, Instituto de F\'isica Interdisciplinar y Sistemas Complejos (CSIC-UIB),
E-07122 Palma de Mallorca, Spain}

\date{\today}


\begin{abstract}

We investigate the relationship between communication and search efficiency in a biological context by proposing
a model of Brownian searchers with long-range pairwise interactions. After a general study of the properties of
the model, we show an application to the particular case of acoustic communication among Mongolian gazelles,
for which data are available, searching for good habitat areas. Using Monte Carlo simulations and density equations,
our results point out that the search is optimal (i.e. the mean first hitting time among searchers is minimum)
at intermediate scales of communication, showing that both an excess
and a lack of information may worsen it.
\end{abstract}

\maketitle

Many living organisms, including bacteria \cite{LiuPassino}, insects, and mammals \cite{dianamonkey,mccomb}
communicate for a variety of reasons including facilitation of social cohesion \cite{cap, pfefferle}, 
defense against predators \cite{predator}, maintenance of territories \cite{stamps,frey2}, and to pool information
 on resource locations when no single individual is sufficiently knowledgeable \cite{bee,hoare,berdahl,simons,torney}.
 Communication among individuals frequently leads to group formation \cite{eftimie}, which often has clear direct benefits such as
 reducing individual vulnerability to predators. Such strategies may, however, also have important incidental benefits.
 For example, an individual that has found a good foraging patch might try to attract conspecifics to reduce its risk of
 predation, but also provides its conspecifics with information on the location of good forage, thus increasing the foraging
 efficiency of those responding to the call.

A variety of mammalian species are known to communicate acoustically over distances of up to several kilometers \cite{mccomb,callhyenas,calllions}, 
but while group formation via vocalizations has been well studied \cite{mccomb,whales,hyenas}, incidental benefits such as increased foraging
efficiency have received little research attention.
In contrast, research on foraging efficiency has focused largely on independent individuals \cite{Benichou2011, vergassola, hein, Viswanathan,viswanathan.nature,bartomeus,mejiamonasterio},
or on comparing foraging behavior across species \cite{klages}.
In addition, recent theoretical work \cite{majumdar} has focused on the statistics of a population of independent random walkers, but an interaction mechanism,
and its influence on search efficiency, has not been thoroughly studied.
To date, very few models have examined the potential effect that long-distance communication \cite{torney2} can have on
movement behaviors and population distributions, and many open questions remain,
particularly on the interrelation between communication and optimal search for resources. 
How can communication facilitate group formation and identification of areas of high quality resources? Does a communication range exist
that optimizes foraging efficiency? To what degree does search efficiency depend on the communication mechanism? Finally, how does communication affect individual space use
in a heterogeneous environment?

In this letter, we address these questions with a general model of random search with two main ingredients: 
resource gradients and long-range communication.
We first introduce a simple theoretical model  (that focuses on large-scale features of the search process 
and does not account for fine-scale details such as collision avoidance, group cohesion or density-dependent diffusivity
\cite{torney,torney2,cates}),
and show how search time changes when 
foragers share information. We then apply a specialized version of the model to the particular case of acoustic communication among
Mongolian gazelle, the dominant wild herbivore in the Eastern steppes of Mongolia. Gazelles must find each
other and relatively small areas of good forage in a vast landscape where sound can travel substantial distances \cite{frey}.
 We aim to explore whether acoustic communication in the Steppe could lead to the formation of observed large aggregations of animals
\cite{olson2009}, and how search efficiency depends on the distances over which calls can be perceived. We wonder
if the frequency of the voice of the gazelles is optimal to communicate in the Steppe, and
if the call length-scales that optimize search in real landscapes are biologically and physically plausible. 
To do this, we couple an individual-based representation of our model with remotely-sensed data on resource quality in the Eastern Steppe.

We consider $N$ particles which undergo a $2-D$ Brownian random walk. Correlated random walks,
often more appropriate to model directional persistence in  animal movement reduce to Brownian motion for large spatiotemporal scales \cite{turchin}.
The movement is biased by the gradients of the landscape quality (local information), and
by the interaction among individuals through a communication mechanism that is
activated when good resources are found, thus providing information on habitat quality in other areas (nonlocal information).
The dynamics of any of the particles $i = 1,..., N$ is 
%%%%%%%%%%%%%%%%%%%%%%%%%%%%%%%%%%%%%%%%%%%%%%%%%%%%%%%%%%%%%%%%%%%%%%%%%%%
\begin{equation}\label{lange}
\dot{\mathbf{r}_i}(t)=B_{g}\nabla g(\mathbf{r}_i)+
B_{C}\nabla S(\mathbf{r}_i)+\mathbf{\eta}_{i}(t),
\end{equation}
%%%%%%%%%%%%%%%%%%%%%%%%%%%%%%%%%%%%%%%%%%%%%%%%%%%%%%%%%%%%%%%%%%%%%%%%%
where $\mathbf{\eta}_{i}(t)$ is a Gaussian white noise term characterized by
$\langle\eta_{i}(t)\rangle=0$, and $\langle\eta_{i}(t)\eta_{j}(t')\rangle=2D\delta_{ij}\delta(t-t')$, 
with  $D$  the diffusion coefficient.
The term $B_{g}\nabla g(\mathbf{r}_i)$ refers to the local search, where $g(\mathbf{r})$ is the
environmental quality function (amount of grass, prey, etc...)
and  $B_{g}$ is the local search bias parameter.
$B_{C}\nabla S(\mathbf{r}_i)$ is the nonlocal 
search term, with 
$B_{C}$ the nonlocal search bias parameter and $S(\mathbf{r}_i)$
is the {\it available information function} of the individual $i$. 
It represents the information arriving at the spatial position of the animal $i$ as a result
of the communication with the rest of the population. 
This term makes the individuals move along the gradients of the information received. 
This is 
a function of the superposition
of pairwise interactions between the individual $i$ and each one of its conspecifics,
\begin{equation}\label{colective}
 S(\mathbf{r}_i)=F\left(\sum_{j=1, j\neq i}^{N} A[g(\mathbf{r}_j)]V(\mathbf{r_i},\mathbf{r}_j)\right).
\end{equation}
$F$ is an arbitrary {\it perception function} that must be set in each application of the model,
$V(\mathbf{r}_i,\mathbf{r}_j)$ is the 
interaction between the receptor particle $i$ depending on its position 
$\mathbf{r}_i$ and the emiting particle fixed at $\mathbf{r}_j$, 
and $A[g(\mathbf{r}_j)]$ is the activation function (typically, a Heaviside function) that
indicates that the individual at $\mathbf{r}_j$ calls the others if it is
in a good habitat. 

From the Langevin equation (\ref{lange}), and following the standard
arguments presented in \cite{Dean1996, Umberto} it is possible to
write an equation for the evolution of the
density of individuals, $\rho (\mathbf{r},t)$. This approach will allow us to
fix the parameters of the problem having a better understanding 
of the role they are playing in the dynamics through a dimensional analysis.
 However, in the case of the large grazing mammals we are going to study later,
it is not very suitable to describe a population as a continuum since 
the number of individuals is not very high and the typical distances among them is large.
Neglecting fluctuations the continuum equation for the density is 
\begin{eqnarray}\label{macro}
 &\frac{\partial\rho(\mathbf{r},t)}{\partial t}=D\nabla^{2}\rho(\mathbf{r},t)+B_{g}\nabla \left[\rho(\mathbf{r},t)\nabla g(\mathbf{r})\right]+&\nonumber \\
&+B_{c}\nabla\left[\rho(\mathbf{r},t)\nabla F \left(\int d\mathbf{r'}\rho(\mathbf{r},t) A[g(\mathbf{r'})]
V(\mathbf{r},\mathbf{r'})\right)\right],&
\end{eqnarray}
which is quite similar to the one derived in \cite{Savelev2005} to study the 
transport of interacting particles on a substrate.

As previously stated, we wish to explore how foraging times are affected 
when individuals share information, but our model could also be generalized to the
case of predators which use prey's signals to locate them, or many other situations where 
animals obtain information from conspecifics.
For the general case, we consider an identity perception function and a Gaussian-like interaction kernel.
Later, to check the robustness of the model to changes in $V$, we will use a physically-motivated power law interaction with an exponential cutoff.
Manipulating its typical range via the standard deviation, $\sigma$, we ask
how the typical communication distance affects the average efficiency of individuals 
searching for targets in space (areas of high-quality forage). We give an answer in terms of
spatial distributions of individuals at long times starting from a random initial condition,
and the mean first arrival time to the targets, $\tau$, as it is done in related works \cite{benichouPRL}.

We begin with Monte Carlo simulations of the individual-based dynamics in eq.~(\ref{lange})
using a square system, $L_{x}=L_{y}=1$, with periodic boundary conditions, and a population
of $N=100$ individuals. We use a theoretical landscape quality function, 
$g(\mathbf{r})$, consisting of three 
non-normalized Gaussian functions, to ensure that $g(\mathbf{r})\in[0,1]$, centered at different spatial points.
The available information function of the individual $i$ depending on its position will be
\begin{equation}\label{perteo}
S(\mathbf{r}_i)=\sum_{j=1, j\neq i}^{N}A[g(\mathbf{r}_j)]\frac{\exp\left(-\frac{(\mathbf{r}_i-\mathbf{r}_j)^{2}}
{2\sigma^{2}}\right)}{2\pi\sigma^{2}},
\end{equation}
where, as mentioned before, $A[g(\mathbf{r})]$ is a theta Heaviside function that
activates the interaction when the quality is over a certain threshold $\kappa$,
$A[g(\mathbf{r})]=\Theta(g(\mathbf{r})-\kappa)$.

\begin{figure}
\centering
\includegraphics[width=0.40\textwidth, clip=true]{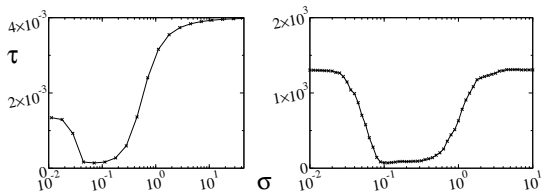}
\caption{Search time using the macroscopic equation (left) with $B_g=0.50$, $B_c=50$, $D=0.75$ and the 
individual based description (right) with $B_g=0.50$, $B_c=0.75$, $D=0.05$. $\kappa=0.85 g_{max}$ in both panels.}
\label{both}
\end{figure}

We observe that the first arrival
time (Fig.~\ref{both} (right)) may be optimized with a 
communication range parameter, $\sigma$, of intermediate scale. 
The number of individuals from which a given animal receives a signal will typically increase with the interaction scale. 
When this scale is too small, individuals receive too little information (no information when $\sigma=0$),  
and thus exhibit low search efficiency (Fig.~\ref{both}). Similarly, interaction scales that are too large lead to 
individuals being overwhelmed with information from all directions, also resulting in inefficient search (Fig.~\ref{both}). In this case, the information
received by any individual is constant over the whole space, 
so that it does not have gradients to follow. Only intermediate communication scales supply
the receiving individual with an optimal amount of information with which to 
efficiently locate the callers and the high-quality habitat areas they occupy.
The same  behavior is also shown by the macroscopic 
equation (\ref{macro}) (left panel in Figure~\ref{both}). 
Now $\tau$ is defined as the time that passes until
half of the population has found a target, that is
$\int_{g(\mathbf{r})\geq\kappa}\rho(\mathbf{r},t)d\mathbf{r} \geq N/2$.
We have integrated the equation (\ref{macro}) in $1-D$ system of length $L=1$, using a single Gaussian patch of 
resources centered at $L/2$ and periodic boundary conditions
for a random initial condition. This is equivalent
to the case of an infinite system with equidistant high quality areas.
We have taken the calling bias as being much stronger than the resource 
bias to make the nonlocal mechanism much more important in the
search process and thus easier to see how the communication 
range parameter affects the search time. The differences between 
the $2-D$ individual-based and the $1-D$ deterministic density equation description, coupled with the parameter choices
(stronger bias in the density equation), explain the different observed time-scales 
in the left and right panels of Figure \ref{both}. The distribution of individuals in the long time limit, 
shows that all the animals end up in good habitats, i.e., in areas where the maxima of the $g$ function occur (not shown).
The values of the threshold $\kappa$, as long as they fall within a reasonable range, 
only change the absolute time scales of the searching process.
%%%%%%%%%%%%%%%%%%%%%%%%%%%%%%%%%%%%%%%%%%%%%%%%%%%%%%%%%%%%%%
%%%%%%%%%%%%%%%%%%%%%%%%%%%%%%%%%%%%%%%%%%%%%%%%%%%%%%%%%%%%%%%%%

Next, we present the application of the model to the Mongolian gazelle ({\it Procapra gutturosa}). 
A detailed analysis of gazelle relocation data has shown that, over the temporal scales relevant 
to searching for resources (days to weeks), Mongolian gazelle movement can be closely approximated
by simple Brownian motion. 
We quantify the habitat quality in the Eastern Steppe of Mongolia using the Normalized
Difference Vegetation Index (NDVI). It is one of the most widely used vegetation quality estimators that can be calculated from satellite imagery, 
and has been already applied to gazelle habitat associations in the Mongolian Steppe \cite{Mueller2008}. 
NDVI is characterized by the function $g_{d}({\bf r})$, a
continuous function taking values between $0$ (no vegetation)  and $1$ (fully vegetated).
As the vegetation at low NDVI is too sparse, and at high NDVI is too mature and indigestible,
gazelles typically seek forage patches characterized by intermediate NDVI values \cite{Mueller2008}.
To make gradients of resources drive the movement of the individuals to
regions with intermediate NDVI values, we apply to the data the following linear transformation:
$g({\bf r})=g_{d}({\bf r})$ if $g_{d}({\bf r})<0.5$, and $g({\bf r})=1-g_{d}({\bf r})$ if $g_{d}({\bf r})>0.5$.
It defines a resources landscape with values between $(0,\ 0.5)$ where $0$ represents both
fully vegetated and no vegetation (i.e., low quality forage).

We study an area of $\sim 23000$~km$^{2}$\cite{area} and assume that the resources remain constant in time during foraging.
It is crucial now to properly choose the perception function in order to 
realistically model the case of gazelles performing {\it acoustic communication}.
It is well known that the sensitivity of the response of the ear does not follow a linear scale, but approximately a logarithmic one. That is 
why the bel and the decibel are quite suitable to describe the acoustic perception of a listener.
Therefore we choose an acoustic perception function of the form
\begin{equation}
 S(\mathbf{r}_i)=10\log_{10}\left(\frac{\sum_{j=1, j\neq i}^{N}A[g(\mathbf{r}_j)]V(\mathbf{r}_i,\mathbf{r}_j)}{I_{0}}\right),
\end{equation}
where the sound calling of $j$, $V(\mathbf{r},\mathbf{r}_j)$, plays the role of a two body interaction potential, and
$I_{0}$ is the low perception threshold (we take the value of a human ear, $I_{0}=10^{-12}~W~m^{-2}$, which
is similar for most other mammals \cite{fletcher}, and in any case, is just a reference value on which our results will not depend). The interaction potential mimicking acoustic communication is
\begin{equation}
 V(\mathbf{r}_i,\mathbf{r}_j)=\frac{P_{0}}{4\pi}\frac{{\rm e}^{-\gamma|\mathbf{r}_i-\mathbf{r}_j|}}{|\mathbf{r}-\mathbf{r}_j|^{2}},
\end{equation}
considering that sound from an acoustic source
attenuates in space mainly due to the atmospheric absorption (exponential term), and the spherical spreading
of the intensity ($4\pi r^{-2}$ contribution), and neglecting secondary effects \cite{naguib}.
 $P_{0}$ may be understood as the power of the sound at a distance
of $1~m$ from the source. The absorption coefficient, $\gamma$ is given by (Stoke's law of sound attenuation \cite{fletcher}) 
$\gamma=\frac{16\pi^{2}\eta\nu^{2}}{3\rho v^{3}}$,      
where $\eta$ is the viscosity of the air, $\rho$ its density, $v$ the propagation velocity of
the acoustic signal (which depends on the 
temperature and the humidity), and $\nu$ its frequency. We work under environmental conditions of $T=20^{o}C$, and
relative humidity of 
$HR=50\%$, which are quite close to the corresponding empirical values for the summer months from the Baruun-Urt (Mongolia) weather station,
averaged over the last 4 years. These values give an absorption coefficient of $\gamma\approx10^{-10}\nu^{2}~m^{-1}$. The inverse
of the absorption coefficient, $\gamma^{-1}$, gives the typical length scale for the communication at each frequency, and thus plays the same
role as the standard deviation, $\sigma$, did in the Gaussian interaction used in the general model. From its functional dependence, different values of the frequency
will modify the value of the absorption coefficient, and consequently, will lead to different communication ranges. Therefore, we will use 
sound frequency, $\nu$, as the control parameter of the interaction range.

From a statistical analysis of GPS data tracking the positions of $36$ gazelles between 2007 and 2011,
we estimate a diffusion constant of $D=74$~km$^{2}$~day$^{-1}$. To give empirically-based values to the bias parameters, we define a 
drift velocity, and based on previous field work \cite{Mueller2008} we set $ v_{drift}=B_{g}\nabla g(\mathbf{r})+B_{c}\nabla S(\mathbf{r})=10~{\rm km~day}^{-1}$.
The local search mechanism is responsible for short-range slow movements, while nonlocal communication gives rise to long and faster movements, and
thus we require $B_{g}\nabla g(\mathbf{r})\ll B_{c}\nabla S(\mathbf{r})$.

\begin{figure}
\begin{center} 
\includegraphics[width=0.35\textwidth]{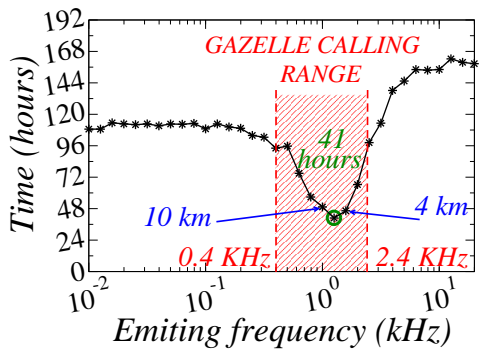}
\caption{(Color online). Mean arrival time for $500$ gazelles (averaged over $50$ realizations with different initial conditions).
Parameter values: $D=74$~km$^{2}$~day$^{-1}$, $B_{g}=2.6\times10^{-3}$~km$^{3}$~day$^{-1}$, $B_{c}=13$~km$^{2}$~day$^{-1}$, $\kappa=0.70g_{max}$.}
\label{kernels}
\end{center}
\end{figure}

\begin{figure}
\begin{center} 
\includegraphics[width=0.30\textwidth]{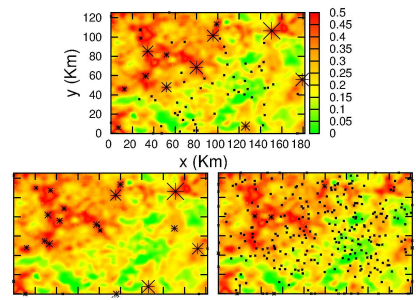}
\caption{(Color online). Spatial distribution of $500$ gazelles after $1$ month (reflecting boundary conditions). $\nu=0.1$~kHz (top), $\nu=1$~kHz (bottom left), $\nu=15.8$~kHz (bottom right).
The size of the star is related to the size of the group at a position. Real data resources landscape.}
\label{100}
\end{center}
\end{figure}

We couple an individual-based model following the dynamics of Eq.~(\ref{lange}), 
with a data-based resources landscape sampled every $500$~m,
and quantify the efficiency of the search for areas of high quality resources in terms of
the mean first arrival time of the population.
We explore the dependence of this metric on the communication length, $\gamma^{-1}$,
or equivalently the frequency, $\nu$ (Figure \ref{kernels}).
Similarly to other species, such as lions \cite{calllions} or hyenas \cite{callhyenas}, the optimal
foraging time ($41$~hours) is obtained for $\gamma^{-1}$ of the order of kilometers (around $6$~km).
This result cannot currently be checked with data. However, switching to frequencies, the optimal search is 
obtained when gazelles communicate at a frequency of $1.25$~kHz,
which lies inside the measured interval of frequencies of the sounds emitted by gazelles, $[0.4, 2.4]$~kHz \cite{freygebler,frey}.
This means that the search is optimal when the receiving individual has an intermediate 
amount of information. A lack of information leads to a slow, inefficient search, while an overabundance of
information makes the individual to get lost in the landscape.
These different regimes are also observed in the long time
spatial distributions (i.e. efficiency of the search in terms of quality) of the Figure \ref{100}.
At intermediate communication scales, $\nu=1$~kHz, (Fig.~\ref{100} bottom left) all of the animals end up in
regions with the best resources, regardless of where they started from. For smaller (Fig.~\ref{100} top) or
larger (Fig.~\ref{100} bottom right) frequencies, some animals are still in low-quality areas at the end of the simulation period.

In summary, our study clarifies some questions on the relationship between 
communication and optimal search for resources. Our key result is that,
in general, intermediate communication distances optimize search
efficiency in terms of time and quality. Individuals are able to find the best quality
resource patches regardless of where they start from,  opening new questions about the distribution 
of individuals in heterogeneous landscapes.
The existence of maximum search efficiency at intermediate communication ranges 
is robust to the choice of functional form of $V(\mathbf{r})$, allowing 
the model to be generalized to many different ways of sharing information.
Another natural extension of the model would be to consider individuals exhibiting L\'evy flight movement behavior. 
This is left for future work, but preliminary results also show the existence of an optimal intermediate communication range.

Communication over intermediate scales results in faster search, and all the individuals form groups in areas of good resources.
While this has obvious advantages in terms of group defense and predator swamping,
it will also lead to rapid degradation of the forage (and thus habitat quality) at those locations. 
This is the problem of foraging influencing the patterns
of vegetation, which will be treated in the future.
Shorter-scale communication implies an almost individual search, which helps preserve local forage quality, 
but has clear disadvantages in terms of group defense strategies. On the other hand,
longer scales lead to the formation of big groups (faster degradation of foraging), 
and animals need more time to join a group,
which has negative consequences against predation. 
Furthermore, acoustic communication scales
significantly larger than the optimal scale for foraging efficiency identified here would be
biologically implausible, even if ultimate group size (and not rate of group formation) was
the most important aspect of an antipredation strategy.
Exploring tradeoffs between group 
defense and individual foraging efficiency in highly dynamic landscapes may be a promising avenue for future research.

R.M-G. is supported by the JAEPredoc program of CSIC. We thank C. H. Fleming
for fruitful discussions and F. Vazquez for fruitful discussions and a critical 
reading of the manuscript. R.M-G. and C.L. acknowledge support from MICINN
(Spain) and FEDER (EU) through Grant No. FIS2007-
60327 FISICOS. J.M.C. and T.M. were supported by a US National Science Foundation grant (ABI 1062411). 

%-----------------------------------------------------------------

\end{document}